
\input phyzzx
\sequentialequations
\overfullrule=0pt
\tolerance=5000
\nopubblock
\twelvepoint

\REF\bekenstein{J. Bekenstein,
{\it Lett. Nuov. Cimento} {\bf 4} (1972) 737,
{\it Phys. Rev.} {\bf D7} (1973) 2333,
{\it Phys. Rev.} {\bf D9} (1974) 3292.}
\REF\hawking{S. Hawking,
{\it Nature (Physical Science)} {\bf 248} (1974) 30,
{\it Comm. Math. Phys.} {\bf 43} (1975).}
\REF\price{R. Price,
{\it Phys. Rev.} {\bf D5} (1972).}
\REF\speculations{L. Susskind,
{\it Some Speculations about Black Hole Entropy in String Theory},
hep-th/9309145 .}
\REF\sen{A. Sen,
{\it Extremal Black Holes and Elementary String States},
hep-th/9504147}
\REF\cvetic{M. Cvetic and D. Youm,
{\it Dyonic BPS Saturated Black Holes of Heterotic String on a
Six Torus},
hep-th/9507090.}
\REF\tseytlin{M. Cvetic and A. Tseytlin,
{\it General class of BPS saturated dyonic black holes as exact
superstring solutions},
hep-th/9510097.}
\REF\susy{R. Kallosh, A. Linde, T. Ortin, A. Peet, and A.V.Proeyen,
{\it Supersymmetry as a cosmic censor},
Phys. Rev. {\bf D}46 (1992) 5278.}
\REF\hairrefs{S. Coleman, J. Preskill, and F. Wilczek,
{\it Growing Hair on Black Holes},
Phys. Rev. Lett. {\bf 67} (1991) 1975; and references therein.}
\REF\duff{M. J. Duff and J. Rahmfeld,
{\it Massive String States as Extreme Black Holes},
Phys. Lett. {\bf B345} (1995) 441.}
\REF\duality{A. Sen,
{\it Strong--Weak Coupling Duality in Four Dimensional String Theory},
Int. J. Mod. Phys. {\bf A9} (1994) 3702;
C. M. Hull and P. K. Townsend,
{\it Unity of Superstring Dualities},
Nucl. Phys. {\bf B438} 109 (1995);
E. Witten,
{\it String Theory in Various Dimensions},
Nucl. Phys. {\bf B443} (1995) 85.}
\REF\peet{A. Peet,
{\it Entropy and Supersymmetry of D Dimensional Extremal Electric
Black Holes versus String States},
hep-th/9506200.}
\REF\callan{C. Callan, J. Maldacena, and A. Peet,
{\it Extremal Black Holes As Fundamental Strings},
hep-th/9510134.}
\REF\horowitz{ G. T. Horowitz and A. A. Tseytlin,
{\it New class of exact solutions in string theory},
Phys. Rev. {\bf D51} 2896 (1995). }
\REF\dabholkar{A. Dabholkar and J. Harvey, {\it Phys. Rev. Lett.} {\bf
63}
(1989) 478; A. Dabholkar, G. Gibbons, J. Harvey, and F. Ruiz-Ruiz,
{\it Nucl. Phys}. {\bf B340} (1990) 33.}
\REF\teitelboim{C. Teitelboim,
{\it Action and Entropy of Extreme and Nonextreme Black Holes},
Phys. Rev. {\bf D51} (1995) 4315, {\it Statistical Thermodynamics
of a Black Hole In terms of Surface Fields}, hep-th/9510180 .}

\line{\hfill }
\line{\hfill PUPT 1576, IASSNS 95/92}
\line{\hfill hep-th/9511064 }
\line{\hfill November 1995}

\titlepage
\title{Internal Structure of Black Holes}

\author{Finn Larsen\foot{Research supported in part by a Danish National
Science Foundation Fellowship.\ \ \ larsen@puhep1.princeton.edu}}
\centerline{{\it Department of Physics }}
\centerline{{\it Joseph Henry Laboratories }}
\centerline{{\it Princeton University }}
\centerline{{\it Princeton, N.J. 08544 }}
\vskip .2cm
\author{Frank Wilczek\foot{Research supported in part by DOE grant
DE-FG02-90ER40542.~~~wilczek@sns.ias.edu}}
\vskip.2cm
\centerline{{\it School of Natural Sciences}}

\centerline{{\it Institute for Advanced Study}}
\centerline{{\it Olden Lane}}
\centerline{{\it Princeton, N.J. 08540}}

\endpage

\abstract{ We present a number of qualitative arguments which strongly suggest
that extremal dyonic black holes in the 4--dimensional low energy,
classical field theory limit of toroidally compactified heterotic
string theory represent largely degenerate classes of states in the
quantum theory.  We propose a simple expression for the full
non--perturbative degeneracy, which contains no free continuous
parameters and reproduces the Bekenstein-Hawking $S={A\over 4G_N}$ in
the large-area limit (with the ${1\over 4}$ arising from
microphysics).  We sketch the elements of a physical picture leading
to this expression: the holes support much hair, such that in counting
it we are led to an effective string theory, and a matching condition
whose solutions give the degeneracy.  }

\endpage

\chapter{Introduction}

The thermodynamic entropy of a black hole
is given by the Bekenstein--Hawking
formula
$$
S_{\rm therm}= {A\over 4\hbar G_N}~,
\eqn\bek
$$
where $A$ is the area of the black hole [\bekenstein -\hawking].
For zero--temperature holes, it is natural to attempt to identify
this entropy with the one we expect from
the fundamental principles of statistical mechanics:
$$
S_{\rm stat}={1\over\hbar}{\rm ln}N
\eqn\stat
$$
where
$N$ is the number of microscopically different states that the black hole
can be in.  Unfortunately the no-hair theorem -- as commonly regarded --
states that the conserved
charges, measurable by observers far away, uniquely specify the black hole
classically, and therefore the statistical mechanical entropy vanishes
classically [\price]. Quantum mechanics may change this, but it seems difficult
to avoid that quantum effects only affect the ${\cal O}(1)$ corrections
to the formulae above which, being of ${\cal O}({1\over\hbar})$, must
have a classical origin\foot{In what follows,
we will set $\hbar=1$.}.

Recently it has been suggested that
string theory might
improve this situation [\speculations ].
In important work Sen [\sen ] offered a concrete
mechanism in the context of
a special class of extremal black holes in toroidally
compactified heterotic string theory.
He noted that
keeping all conserved charges fixed there remain additional
degeneracies among string states with the quantum numbers of the black
hole,
associated with freedom in the choice of transverse
oscillation.
The holes considered by Sen have zero area, and therefore zero
entropy, classically, so they offer a limited perspective on the
problem mentioned above.
Cvetic and various collaborators [\cvetic -\tseytlin ] have
generalized the solutions used by Sen to a class including extremal
holes with non-zero area, and these provide an appropriate arena to
consider the general problem.  They retain  important
technical features of the special case: in particular,
they are BPS saturated
states with $N=1$ (but not $N=2$) supersymmetry [\susy ].

Here we consider the structure of these black holes, and attempt to
identify the required degrees of freedom.
The essence of the matter is that the
no-hair ``theorem'' is not a law of Nature -- indeed, isolated
exceptions
have been noted on several occasions [\hairrefs ].
In the field theories that arise as
the low-energy limit of string theory, one finds that they fail in a
spectacular way: appropriate black holes exhibit a string theory's
worth
of degeneracy, with a string tension renormalized from the microscopic
value in a way that depends on the properties of the hole.
This classical
hair provides a microphysical
grounding for the Bekenstein-Hawking entropy.

\chapter{Classical Black Holes and Suggestions of a State-Count}

We now briefly review the properties of the black holes of interest,
and take special note of some very special properties of the area,
which already suggest a state-counting interpretation.

We consider the low energy limit of
heterotic string theory compactified on a generic torus.
In Einstein metric the effective $4$--dimensional action is
$$\eqalign{
L= {1\over 16\pi G_N}&\int d^4 x \sqrt{-g}
[R_g-{1\over 2}\partial^\mu\phi\partial_\mu\phi
-{1\over 12}e^{-2\phi}
H_{\mu\nu\rho}H^{\mu\nu\rho}\cr
&-G_N e^{-\phi}F_{\mu\nu}^{(a)}
(LML)_{ab}F^{(b)\mu\nu}
+{1\over 8}{\rm Tr}(\partial^\mu ML\partial_\mu ML)
+{\rm fermion~terms}]\cr}
\eqn\action
$$
The fermion terms are determined by supersymmetry from the ones given.
$
H^{\mu\nu\rho}= -{1\over\sqrt{-g}}e^{2\phi} \epsilon^{\mu\nu\rho\sigma}
\partial_\sigma\Psi$ defines the axion $\Psi$ which we will be assume to
vanish, with no essential loss of generality.
The vacuum expectation value of the dilaton $\phi$ is related to the string
coupling through $\langle e^\phi\rangle =e^{\phi_\infty}=g^2_{\rm st}$.
The dimensionful string parameter $\alpha^\prime$ has been
scaled out using $G_N ={\alpha^\prime g^2_{\rm st}\over 4}$.
$F^{(a)}_{\mu\nu}$ represents $28$ $U(1)$--gauge fields. The matrix
$M$ parameterizes the metric of the compactified dimensions and
other background fields. It satisfies
$$
MLM^T =L, ~~~ M^T=M, ~~~ L=
\left(\matrix
{0 & I_6 & 0 \cr
 I_6 & 0 & 0 \cr
 0 & 0 & -I_{16}\cr}\right)
\eqn\Mmatrix
$$

Now consider extremal black hole solutions of this classical field theory.
The asymptotic behavior of $F_{\mu\nu}^{(a)}$ at infinity identifies $28$
electric charges $\vec{Q}$ and $28$ magnetic charges $\vec{P}$.
For the supersymmetric black holes that we are interested in, the full solution
depends only on these charges. Specifically, the dependence
on $g_{\rm st}$ and the background $M_\infty$ is determined by
symmetries of the Lagrangean. Moreover, taking
$\Omega\in{\cal O}(6,22)$  ({\it i.e.} $\Omega^T L\Omega = L$),
the action is invariant under the T-duality transformation
$M\rightarrow \Omega M\Omega^T$ and
$A_\mu^{(a)}\rightarrow \Omega_{ab}A_\mu^{(b)}$ with the other fields
kept fixed. Therefore the solution depends only on T--invariant
combinations of the charges.
For vanishing axion field
it can be shown [\cvetic ] that supersymmetry
imposes the constraints
$\vec{Q}^T [(LML)_\infty\pm L]\vec{P}=0$. The remaining
T--invariants are
$$
Q_{R,L}= |~\vec{Q}^T [(LML)_\infty\pm L] \vec{Q} ~|^{1\over 2}~,
\eqn\Tinvs
$$
and similarly for $P$. The extremal black hole
solutions are explicitly known
[\cvetic ] in terms of the T--invariants
but we will only need the relations
$$
\eqalign{
G_N M^2_{\rm ADM} &={1\over 2}[{1\over g_{\rm st}}P_R+
g_{\rm st}({Q_R\over g^2_{\rm st}})]^2 \cr
S_{\rm therm} &\equiv
{A\over 4G_N}=2\pi\sqrt{(P_R^2-P_L^2)(Q_R^2-Q_L^2)\over g^4_{\rm st}}\cr
}
\eqn\bhprops
$$
Regular solutions satisfy $P_R\geq P_L$ and $Q_R\geq Q_L$.

Due to the normalization of the kinetic energy of the gauge field it
is $e^{-\phi}F_{\mu\nu}^{(a)}$ rather than $F^{(a)}_{\mu\nu}$ that
leads to a conserved charge. The Bianchi identity, on the other hand,
is the standard one so
the conserved charges are $\vec{Q}\over g^2_{\rm st}$ and $\vec{P}$ .
The mass relation has been extensively studied recently [\duff--\duality ].
Keeping $G_N$ and the conserved charges fixed, the
black holes with vanishing magnetic charge are light for small
coupling $g_{\rm st}$, and we expect them to be described
by elementary string states.
In contrast, all magnetically charged black holes are
heavy for small coupling, and we expect them to be described
by non--perturbative, or collective, states.

The entropy as defined above is simply
a geometric parameter of the classical solution.
Yet it exhibits several remarkable features, which are not at all
implicit in its definition, and which strongly suggest a
state-counting
interpretation.
Specifically, unlike $P_R$ and $P_L$ separately
the combination $P_R^2-P_L^2$
is independent of the moduli $M_\infty$. The same is true for $Q$, and
hence for the ``entropy'' $S_{\rm therm}$.
This is reminiscent of the classical
independence of the phase space volume under canonical transformations
-- or, more directly in our context, the expected independence of the
{\it number\/} of states with given conserved quantum numbers at
infinity on adiabatic changes of underlying parameters.
Also, when expressed in terms
of conserved charges the entropy is independent
of the string coupling.
Again, this is consistent with one's  expectation that
parameters of the states (such as
the mass), but not their
total number,  may depend on the coupling constant.
The independence of coupling constant, more rigorously grounded in the
principle of BPS saturation, is also welcome on technical
grounds: it can be convenient to count the states for an arbitrarily weakly
coupled theory.

\chapter{Heuristic Counting of String States}

In classical Einstein-Maxwell theory
an extremal black hole configuration is described uniquely in
terms of the electric and magnetic charges. In string theory
such a state may have some degeneracy $d(\vec{Q},\vec{P})$.
This quantity is meaningful even in the interacting theory where
the internal structure of the states may be complicated.  In this section
we will determine the degeneracy using general principles, and a natural
hypothesis.

First, string theory respects T-duality, so
{\it a priori} the degeneracy may depend on the T--invariants
$P_{R,L}$ and $Q_{R,L}$ as well as the string coupling $g_{\rm st}$.
However, the dependence on continuous parameters $M_\infty$
(implicit in $P_{R,L}$ and $Q_{R,L}$) and $g_{\rm st}$ must respect that
the degeneracy is an integer.  If we assume that the degeneracy is
an analytic function of its variables, the most we can reasonably
hope is that it
takes on integer values at discrete points.
Thus plausibly the only independent parameters that can appear as
arguments of $d$ are
the quantized charges ${P_R^2-P_L^2\over 2}$ and
${Q_R^2-Q^2_L\over 2g^4_{\rm st}}$. (Our normalization is such that these
two variables take on all integer values.)  Next, let us take into account
the symmetry under S--duality, which is conjectured to
be an exact quantum symmetry of the full non--perturbative string theory
[\duality ]. In the absence of an axion, S--duality interchanges
$\vec{P}\leftrightarrow{1\over g^2_{\rm st}}\vec{Q}$ and
$g_{\rm st}\leftrightarrow{1\over g_{\rm st}}$.
Thus the most general expression consistent with symmetries
takes the form
$$
d(\vec{Q},\vec{P})=
F(~{Q_R^2-Q^2_L\over 2g^4_{\rm st}}+{P_R^2-P^2_L\over 2}~,~~
{(Q_R^2-Q^2_L)(P_R^2-P^2_L)\over 4g^4_{\rm st}}~)~.
\eqn\degfunc
$$

For black holes with vanishing magnetic charge we can calculate the
degeneracy in perturbation theory, following Sen [\sen ]:
the mass of a string state is given in terms of its charges
by
$$
G_N M^2_{\rm ADM}= g^2_{\rm st} ({1\over 2g^4_{\rm st}}Q^2_R +N_R-{1\over 2})=
g^2_{\rm st} ({1\over 2g^4_{\rm st}}Q^2_L +N_L-1)
\eqn\massform
$$
For extremal states the supersymmetry algebra determines
$N_R={1\over 2}$, so we recover the mass formula from the black holes in
the previous section. This is the identification between black holes
and string states found in [\duff ].
The second equality is the level matching condition, which
enforces reparameterization invariance along the string.
For a given set of charges the oscillator level $N_L$ is given by \massform\
but there any many ways that oscillations can
be distributed on individual string modes. The degeneracy is
$$
d(\vec{Q}, \vec{P}=0)\equiv f(N_L)
\simeq e^{4\pi\sqrt{N_L}}~.
\eqn\degen
$$
The function $f(N_L )$ is known, but complicated, so we use its
leading term for large $N_L$. The statistical entropy is
$$
S_{\rm stat}= {\rm ln} f(N_L)\simeq
4\pi\sqrt{{Q_R^2-Q^2_L\over 2g^4_{\rm st}}+1}
\eqn\matchtoent
$$
For vanishing magnetic charge the corresponding
thermodynamic entropy calculated in the previous section is zero.
However, it has been argued that the thermodynamic entropy receives
corrections of a form consistent with the statistical entropy above
[\sen, \peet ].

The perturbative calculation determines the dependence of the general
degeneracy on ${Q_R^2-Q^2_L\over 2g^4_{\rm st}}+{P_R^2-P^2_L\over 2}$.
Unfortunately the other variable
${(Q_R^2-Q^2_L)(P_R^2-P^2_L)\over 4g^4_{\rm st}}$ vanishes for zero magnetic
charge, so
the perturbative calculation teaches us nothing about its significance.
Specifically, the general relation might, {\it a priori}, be independent of
this parameter.  We shall now argue  this cannot be true.
Consider the regime where at least one of the charges is large.
Consider the number of states consistent with asymptotic charges
$\vec{P_1}+\vec{P_2}$ and $\vec{Q_1}+\vec{Q_2}$.  A subclass of such
states
consists of well separated subsystems each carrying part
of the charge. We therefore expect
$$
d(\vec{Q_1}+\vec{Q_2},\vec{P_1}+\vec{P_2})\geq
d(\vec{Q_1},\vec{P_1})d(\vec{Q_2},\vec{P_2})~.
\eqn\cluster
$$
It requires the degeneracies to be at least exponential
in electric as well as magnetic charges, and is inconsistent with
$F$ being independent of its second argument.

As an {\it ansatz} let us assume the existence of
a generalized matching condition
equating $N_L$ to some function of $\vec{Q}$ and $\vec{P}$.
In view of the clustering property, this function can not
simply be the sum of terms depending only on
$\vec{Q}$ and $\vec{P}$ separately.
As the next simplest choice let us consider
a product of terms depending on $\vec{Q}$ and $\vec{P}$ respectively.
In fact, taking into account the known result
for $P=0$, there is a {\it unique} factorized matching condition
$$
N_L ~=~  ({Q_R^2 - Q_L^2 \over 2g^4_{\rm st} } +1) ~~
( {P_R^2 - P_L^2 \over 2 } + 1)~,
\eqn\levels
$$
It is an independent consistency check, although admittedly weak,
that this relation equates integers. Thus we are led to
$$
S_{\rm stat}={\rm ln}~f(N_L)\simeq
4\pi\sqrt{({Q_R^2-Q^2_L\over 2g^4_{\rm st}}+1)({1\over 2}(P_R^2-P^2_L)+1)}~.
\eqn\answer
$$
Remarkably, this expression reproduces  $S_{\rm therm}$ for large charges.
In particular, it gives the numerical coefficient in front
of the area correctly.
In its entirety it is a corrected form of the thermodynamic entropy
formula, that interpolates between Sen's perturbative result and
the classical expression.

In the following section we shall consider the problem from a more
microscopic perspective, and argue that the existence of a
matching condition in factorized form
is not only esthetically pleasing but also physically necessary.
Taken together with the earlier arguments,
this singles out \answer\ uniquely.
Let us note the
microphysical origin of the
various coefficients.
The 1's that quantum correct the electric and magnetic charges
arise from world sheet zero point energy. The coefficient in front of
the square root is actually $2\sqrt{{\pi^2\over 6}\times 24}$ which
is related to the world sheet entropy $\pi^2\over 6$
of the 24 transverse oscillators of the left
movers of the heterotic string.

\chapter{Black Hole Hair}

The static black hole solution is specified uniquely by $\vec{Q}$, $\vec{P}$,
and the extremeality condition [\cvetic ]. In a generic theory one
would expect that time-dependent perturbations are
exponentially damped outside the horizon --
black holes have no hair [\price ].
However string theory dictates additional internal dimensions
and a very special
matter content so, perhaps surprisingly,
the black hole allows carefully chosen
perturbations. To exhibit these in a simple form we use the string metric
$G_{\mu\nu}=e^{\phi}g_{\mu\nu}$.
We also single out one of the
compactified directions, $x^{9}$ say, and let the two independent
electric charges of the background
be associated with the Kaluza--Klein field $G_{9t}$
and the antisymmetric field $B_{9t}$, with the other internal
dimensions carrying no electric charge.
The modes
$$\eqalign{
G_{ui}(u,r)&=\partial_u X^i (u)~C(r)=B_{ui}(u,r)~;~~~i=1,...,8 \cr
A^{J}_{u}(u,r)&=\partial_u X^{J+8}(u)~C(r)~;~~~J=1,..., 16 \cr}
\eqn\modes
$$
where $u=x^{9}-t$ and $v=x^{9}+t$, propagate
freely in this background.  The functions $X^I(u)$ ($I=1,...,24$) are
{\it arbitrary},  except that they must
respect the periodicity of the compactified coordinate $x^{9}$.
The profile function is
$C(r)={g^2_{\rm st}\over r+\sqrt{\alpha^\prime /2}(Q_R-Q_L)}$.
Using the techniques of [\tseytlin, \callan --\horowitz ],
each of the perturbations can be extended to exact solutions,
respecting supersymmetry. This even holds at the level of the exact
conformal field theory,
which is the string-theoretic analogue of a classical
state . There is a simple physical interpretation of this
hair: the translational invariance of the underlying 10--dimensional
theory is broken by the location of the black hole. After gauge--fixing
this leads to 8 Nambu-Goldstone modes $\partial_u X^i$.
Analogously the internal
16--torus leads to the remaining $\partial_u X^I$.
As we shall explain below, the background allows
only left moving perturbations, and no dependence on internal
coordinates other than $x^{9}$.

In deriving equations of motion from the action we are instructed to
consider variations of the fields, keeping them fixed on the
boundaries.
Besides the bulk Euler-Lagrange equations, this implies regularity
conditions at real or coordinate singularities.  In the present
context, by considering the coefficient in variations of
$\partial_r G_{vv}$ near the horizon $r=0$ we are led --
after a substantial calculation --
to a matching condition for the left movers
$$
\sum_{I=1}^{24}  [\partial_u X^I (u) ]^2 = {1\over g^4_{\rm st}}
(Q_R^2-Q_L^2) ~.
\eqn\formofmatching
$$
Right movers are prohibited by the similar argument, with $u$ replaced by $v$.
Since the electric charge is in the $x^9$ direction,
the corresponding condition in directions $\bar{u}=x^i-t$ (with
$i=4,...,8$) forbids oscillation. This is why we only find a string's
worth of oscillators, rather than those characteristic of
some higher p--brane.  These matching conditions should be considered
as a mathematical embodiment of the requirement that there are no
physical sources at the horizon.
For electric black holes Callan {\it et. al.}
[\callan ] obtain a generalized
matching condition of the same form by demanding a form of cosmic censorship.
Closely related matching conditions occur in other contexts
[\dabholkar ].

The low energy dynamics of black holes is governed by that of the
0--modes, which in turn realize an effective heterotic string theory with the
right movers in their ground states.
To count states we must find the normalization of
the functions $X^I (u)$; more precisely,
the number of states in the semi-classical approximation is determined
by the volume of classical phase space.
The momentum conjugate to the collective coordinate
$X^I$, denoted $\Pi^I$,
is given in light-cone quantization (with $v$ as the ``time'')
by the variational derivative of the Lagrangean with respect
to $\partial_v X^I$. Note that while $\partial_v X^I =0 $ on any solution
of the equation of motion, the momenta need not vanish.
The result must be of the form $\Pi^I = {T\over 2}\partial_u X^I$, which
after quantization leads to the generalized matching condition
$$
N_L = (\pi\alpha^\prime T) { Q^2_R-Q^2_L \over g^4_{\rm st} }
\eqn\whatever
$$
up to a normal ordering constant.
$N_L$ is the oscillator level of the $24$ 0--modes.

Due to a scaling symmetry of the classical action,
the purely electrical black hole depends only on the parameters
$\sqrt{\alpha^\prime}{Q_{R,L}\over r}$.
In the weak coupling limit,
keeping the conserved charges ${Q_{R,L}\over g^2_{\rm st}}$ fixed,
the black hole reduces to flat space.  In this limit, assuming that
states of the effective string theory are to be identified with
elementary string states with the same quantum numbers, we must put
$T={1\over 2\pi\alpha^\prime}$.
It is a non-trivial coincidence,
that the properly normalized classical regularity condition for
0-modes in an electric background agrees with the elementary string
matching condition.

Adding magnetic
charges, the black hole background depends on
$\sqrt{\alpha^\prime}{P_{R,L}\over r}$.
In the weak coupling limit
the tension depends on the magnetic, but not the electric charges.
This is the origin of the factorization of the matching condition.
Given this factorization, the general considerations
of the previous section determines the degeneracy
$d(\vec{Q}, \vec{P})$ {\it uniquely}, in the form previously mentioned.

Given the explicit form of the 0--modes
the tension, determined indirectly above,
can in principle be calculated directly.  Since the
required Hamiltonian formalism is both complex and
notoriously subtle due to the need for careful attention to boundary
terms, the following remarks should be regarded as
indicative rather than definitive.
We find for the bulk tension
$$
T = {1\over 2\pi\alpha^\prime}\int d^4 x \sqrt{-G}e^{-\Phi}[G^{uv}G^{ij}
\partial_i C\partial_j C ]
\eqn\tensionexp
$$
where the integral is over three spatial directions and the
internal dimension carrying the magnetic charges.
In the purely electric case the tension is finite but
vanishes as $g_{\rm st}\rightarrow 0$, so we fail to recover
the flat space result.  This
is not unexpected from a physical point of view,
because the action from which we derived the explicit form of the
solution is not uniformly accurate.  At scales of order
${\cal O}(\sqrt{\alpha^\prime})$ close to the horizon we
cannot trust the form of the metric to be inserted into \tensionexp.
Fortunately the weak coupling limit determines
the true tension, as we have seen.

In the generic case, one may take the electric charge smoothly to zero as
the coupling becomes weak at infinity.
The remaining magnetic soliton
is controlled by an effective coupling which becomes of order unity at a large
distance from the horizon. We assume -- motivated by duality -- that
universal physics governs the interior strong coupling region.
Concretely, we suppose the
expression in brackets in \tensionexp\ is universal and
put $\sqrt{-G} e^{-\Phi}\simeq {1\over 2}(P^2_R-P^2_L)$ -- its
value where the coupling become strong.
The matching condition then emerges in the anticipated, factorized, S-dual
form.

While our technical implementation
certainly leaves much room for improvement,
the emergent physical picture seems compelling:
the dyon has an aspect that is electric and
singular at short distances, cut off in a universal manner; and
also a soft solitonic, magnetic
aspect which gives it a classical size, so that this universal
physics is repeated over a large volume.  Let us emphasize again that this
qualitative picture, combined with the weak coupling limit, allows one
to determine the degeneracy {\it quantitatively}.

\chapter{Discussion}

The idea that there is a very large density of states arising from the
normalization of a degree of freedom associated with fluctuations of
the horizon has also been proposed on apparently very different
grounds, with no explicit reference to string theory but crucially
involving a (notional) general covariant cut-off in quantum gravity,
by Teitelboim [\teitelboim ].  We suspect there are deep connections between
these points of view.

The black holes considered above are not as exotic as they
might appear at first sight.
For example, they include as a special case
the extremal Reissner--Nordstr\o m black holes [\duff ]. Those correspond
to $Q_L=P_L=0$ and $Q_R=P_R={g_{\rm st}\over\sqrt{2}}Q_{\rm el}$.  This
identification is possible because
the dilaton decouples in this limit.
The additional structure in our construction
serves to embed standard electromagnetism
in a theory that is supersymmetric,
to avoid quantum corrections. It also allows a precise connection
with string theory.
Based on the adiabatic invariance of state
number
one would expect that, to the extent that all string theories are
continuously connected,
details of the embedding,
the class of string theory, and the compactification are unimportant.
We
have seen this to be the case, for a restricted but non-trivial
class of moduli.   Because our picture of black hole internal
structure -- classical hair -- is so simple and concrete, we also
expect that it applies to all black holes, extremal or not.

{\bf acknowledgments}

We would like to thank C. Callan, D. Gross, J. Maldacena, and H. Verlinde
for discussions.
We have also benefited from suggestions by M. Cvetic,
A. Peet, E. Sharpe, and L. Susskind.

\refout
\bye